\documentclass[twoside,twocolumn,9pt]{article}
\usepackage{extsizes}
\usepackage[super,sort&compress,comma]{natbib} 
\usepackage[version=3]{mhchem}
\usepackage[left=1.5cm, right=1.5cm, top=0.0cm, bottom=5cm]{geometry}
\usepackage{balance}
\usepackage{times,mathptmx}
\usepackage{sectsty}
\usepackage{graphicx} 
\usepackage{lastpage}
\usepackage[format=plain,justification=justified,singlelinecheck=false,font={stretch=1.125,small,sf},labelfont=bf,labelsep=space]{caption}
\usepackage{float}
\usepackage{fancyhdr}
\usepackage{fnpos}
\usepackage[english]{babel}
\addto{\captionsenglish}{%
  
}
\usepackage{array}
\usepackage{droidsans}
\usepackage{charter}
\usepackage[T1]{fontenc}
\usepackage[usenames,dvipsnames]{xcolor}
\usepackage{setspace}
\usepackage[compact]{titlesec}
\usepackage{hyperref}

\DeclareMathOperator*{\argmin}{arg\,min}

\definecolor{cream}{RGB}{222,217,201}

\begin{document}
\pagestyle{fancy}
\thispagestyle{plain}
\fancypagestyle{plain}{

\fancyhead[L]{\hspace{0cm}\vspace{1.5cm}}
\fancyhead[R]{\hspace{0cm}\vspace{1.7cm}}
\renewcommand{\headrulewidth}{0pt}
}

\makeFNbottom
\makeatletter
\renewcommand\LARGE{\@setfontsize\LARGE{15pt}{17}}
\renewcommand\Large{\@setfontsize\Large{12pt}{14}}
\renewcommand\large{\@setfontsize\large{10pt}{12}}
\renewcommand\footnotesize{\@setfontsize\footnotesize{7pt}{10}}
\makeatother

\renewcommand{\thefootnote}{\fnsymbol{footnote}}
\renewcommand\footnoterule{\vspace*{1pt}%
\color{cream}\hrule width 3.5in height 0.4pt \color{black}\vspace*{5pt}} 
\setcounter{secnumdepth}{5}

\makeatletter 
\renewcommand\@biblabel[1]{#1}            
\renewcommand\@makefntext[1]%
{\noindent\makebox[0pt][r]{\@thefnmark\,}#1}
\makeatother 
\renewcommand{\figurename}{\small{Fig.}~}
\sectionfont{\sffamily\Large}
\subsectionfont{\normalsize}
\subsubsectionfont{\bf}
\setstretch{1.125} 
\setlength{\skip\footins}{0.8cm}
\setlength{\footnotesep}{0.25cm}
\setlength{\jot}{10pt}
\titlespacing*{\section}{0pt}{4pt}{4pt}
\titlespacing*{\subsection}{0pt}{15pt}{1pt}

\fancyfoot{}
\fancyhead{}
\renewcommand{\headrulewidth}{0pt} 
\renewcommand{\footrulewidth}{0pt}
\setlength{\arrayrulewidth}{1pt}
\setlength{\columnsep}{6.5mm}
\setlength\bibsep{1pt}

\makeatletter 
\newlength{\figrulesep} 
\setlength{\figrulesep}{0.5\textfloatsep} 

\newcommand{\topfigrule}{\vspace*{-1pt}%
\noindent{\color{cream}\rule[-\figrulesep]{\columnwidth}{1.5pt}} }

\newcommand{\botfigrule}{\vspace*{-2pt}%
\noindent{\color{cream}\rule[\figrulesep]{\columnwidth}{1.5pt}} }

\newcommand{\dblfigrule}{\vspace*{-1pt}%
\noindent{\color{cream}\rule[-\figrulesep]{\textwidth}{1.5pt}} }

\makeatother

\twocolumn[
  \begin{@twocolumnfalse}
\vspace{3cm}
\sffamily
\begin{tabular}{m{4.5cm} p{13.5cm} }

 & \noindent\LARGE{\textbf{Spatial Damage Characterization in Self-Sensing Materials via Neural Network-Aided Electrical Impedance Tomography: A Computational Study}} \\
\vspace{0.3cm} & \vspace{0.3cm} \\

 & \noindent\large{Lang Zhao,\textit{$^{a}$} Tyler Tallman,\textit{$^{b}$} and Guang Lin\textit{$^{c}$}$^{\ast}$} \\

 & \noindent\normalsize{Continuous structural health monitoring (SHM) and integrated nondestructive evaluation (NDE) are important for ensuring the safe operation of high-risk engineering structures.  Recently, piezoresistive nanocomposite materials have received much attention for SHM and NDE. These materials are self-sensing because their electrical conductivity changes in response to deformation and damage. Combined with electrical impedance tomography (EIT), it is possible to map deleterious effects. However, EIT suffers from important limitations -- it is computationally expensive, provides indistinct information on damage shape, and can miss multiple damages if they are close together. In this article we apply a novel neural network approach to quantify damage metrics such as size, number, and location from EIT data. This network is trained using a simulation routine calibrated to experimental data for a piezoresistive carbon nanofiber-modified epoxy. Our results show that the network can predict the number of damages with 99.2\% accuracy, quantify damage size with respect to the averaged radius at an average of 2.46\% error, and quantify damage position with respect to the domain length at an average of 0.89\% error. These results are an important first step in translating the combination of self-sensing materials and EIT to real-world SHM and NDE.}

\end{tabular}

 \end{@twocolumnfalse} \vspace{0.6cm}

  ]

\renewcommand*\rmdefault{bch}\normalfont\upshape
\rmfamily
\section*{}
\vspace{-1cm}


\footnotetext{\textit{$^{a}$~School of Mechanical Engineering, Purdue University, West Lafayette, IN 47907, USA}}
\footnotetext{\textit{$^{b}$~School of Aeronautics and Astronautics, Purdue University, West Lafayette, IN 47907, USA}}
\footnotetext{\textit{$^{c}$~Department of Mathematics, School of Mechanical Engineering, Department of Statistics (Courtesy), Department of Earth, Atmospheric, and Planetary Sciences (Courtesy), Purdue University, West Lafayette, IN 47907, USA}}
\footnotetext{~Corresponding author, Fax: 765 494 0548; Tel: 765 494 1965; E-mail: Guanglin@purdue.edu}








\section{Introduction}

Structural health monitoring (SHM) and integrated nondestructive evaluation (NDE) are promising methods of ensuring the safety and reducing the maintenance costs associated with potentially high-risk and high-value aerospace, civil, and energy engineering structures. \cite{farrar2007introduction} Recently, so-called \emph{self-sensing} materials have received much attention for application in SHM and integrated NDE. These materials are engineered to elicit a property change in response to environmental factors. For mechanical self-sensing (i.e. the detection of deformation and damage), the piezoresistive effect has much potential. Piezoresistivity refers to a material having deformation and damage dependent electrical conductivity. Piezoresistive-based self-sensing is very appealing for SHM and integrated NDE because low-cost, non-invasive, and easily multiplexed electrical measurements can be used to monitor the mechanical state of the material or structure. 

Piezoresistivity is typically achieved by modifying a non-conductive matrix with highly conductive fillers. When sufficiently many conductivity fillers are added, an electrically connected or percolated filler network forms thereby allowing for electrical transport through the insulating matrix. Deformations which change the connectedness of the fillers or damage which severs the connection between fillers manifest as a conductivity change. The concentration of fillers needed to achieve percolation is closely tied with filler aspect ratio -- very long and thin fillers percolate at much lower filler weight fractions. Because it is desirable to not modify the host matrix with excessively many fillers (potentially adversely affecting the load-carrying capability of the material), ultra-high aspect ratio fillers such as carbon nanotubes (CNTs) and carbon nanofibers (CNFs) have been thoroughly studied for use in self-sensing materials. For example, nanofiller modification has been explored for imparting piezoresistive self-sensing to fiber-reinforced polymers, \cite{thostenson2006carbon} polymer sensing skins, \cite{loh2009carbon} ceramics, \cite{inam2014structural} and even cementitious materials. \cite{d2016investigations} 

Electrical impedance tomography (EIT) is a natural complement for SHM in piezoresistive materials. Quite succinctly, EIT is a process by which the internal conductivity distribution of a domain can be spatially mapped via a series of voltage-current measurements taken at the boundary of the domain. EIT therefore allows us to detect and spatially map conductivity-changing artifacts such as deformations in nanofiller-modified polymers, \cite{tallman2015tactile} impact damage in nanofiller-modified glass fiber/epoxy composites \cite{thomas2019damage} and sensing skins, \cite{loyola2013spatial} and damage to cement with CNT-modified aggregate interfaces. \cite{gupta2017self} Beyond nanofiller-modified materials, EIT has also been used for detecting and shaping damage in cementitious materials with silver or copper-based conductive coatings, \cite{smyl2018detection} metal particle-reinforced cement, \cite{nayak2019spatial} and carbon fiber-reinforced composites. \cite{cagan2020impact} Because of this diversity of material systems and ability to shape and localize damage, the combination of piezoresistive self-sensing and conductivity mapping via EIT seemingly has much potential for SHM and integrated NDE.

\begin{figure}[h]
    \centering
    \includegraphics[width=8cm]{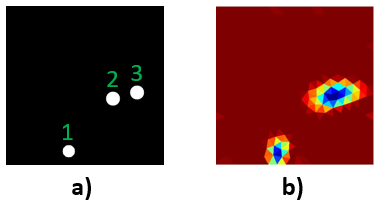}
    \caption{Demonstration of important limitations of EIT for damage detection. a) Three holes are simulated. Two of the holes are quite close together. b) While the EIT reconstruction does capture artifacts collocating with the holes, the image is blurry making it difficult to extract meaningful diagnostic information such as hole size. Worse, EIT blurs together holes 2 and 3. This demonstrates an important limitation of EIT -- it can fail to differentiate between close damages.}
    \label{Fig:EIT_limitation}
\end{figure}

Despite this potential, EIT has some important limitations. First, depending on its formulation, it can be computationally expensive and therefore slow to produce an image (e.g. nonlinear difference imaging \cite{liu2015nonlinear}). Second, EIT images are blurry and indistinct. This can make definitive damage characterization difficult. This limitation is a consequence of the underlying physics of EIT being diffusion-based. And third, because of the diffuse nature of EIT, it can also miss multiple damages that are close together such as closely spaced holes by reproducing a single, large blob-like conductivity change that covers both damages. These second and third limitations -- a lack of distinct shaping and indistinguishability of closely spaced damages -- are readily apparent in Figure \ref{Fig:EIT_limitation}. In this figure, EIT blurs together two closely spaced holes. Manual interpretation of EIT for prognostic purposes may therefore mistakenly assume that there are only two damages when there are, in fact, three. Furthermore, manual interpretation would be unable to provide precise characterization of these damages beyond occuring in roughly the same location as the conductivity change and being of roughly the same shape. Overcoming these limitations is important because real-time damage quantification (i.e. knowing the precise damage location, size, and shape) is essential to effecting meaningful diagnostics.

In light of the preceding discussion, we herein present a novel method of integrating machine learning via neural networks with the EIT problem for damage detection, localization, and quantification. Indeed, the application of machine learning and neural networks to EIT has already received some attention particularly in the context of medical imaging. \cite{hu2019image, hamilton2018deep, huang2019improved} However, there are some important differences in the work herein presented. First, we apply network-supplemented EIT for damage detection in an engineered material system rather than for physiological imaging. And second, we train the network to \emph{parameterize} the damage. That is, some \emph{a priori} information on the damage is provided to the network such that it can very accurately predict damage parameters including location, size, and number in a piezoresistive material with accuracy that is unapproachable by standard EIT. 

In the fourthcoming we will first describe the material system by which our computational study was calibrated. Next, we will summarize the standard EIT formulation. Then, we will describe two approaches to training the neural network -- via EIT conductivity maps (i.e. image data) and via EIT-predicted conductivity change vectors (i.e. numerical data). It will be shown that this is an important distinction which substantially affects the accuracy of the network. And lastly, we computationally validate the network using an experimentally calibrated simulation routine. Summarily, the intent of this work is to explore whether or not EIT-generated conductivity data (i.e. in the form of maps and vectors) can be integrated with existing machine learning architectures for spatial damage characterization.

\section{CNF/epoxy}

The networks herein utilized were computationally trained using a model calibrated to experimental data for a CNF/epoxy nanocomposite. For the sake of completeness, the manufacturing procedure is briefly outlined here. However, interested readers are directed to the original work by Tallman and Hassan \cite{tallman2019network} for a more expansive discussion on how the material was manufactured and its electrical properties were measured. To produce this material, CNFs (Pyrograf III PR-24-XT-HHT, Applied Sciences) were dispersed into an uncured liquid epoxy resin (Fibre Glast 2000) using a combination of high-energy mixing, bath sonication, a viscosity thinning agent, and chemical surfactant (Triton X-100, bioWORLD). Procedurally, desired amounts of CNFs were first weighed and added to the liquid epoxy resin along with acetone at an acetone-to-resin volume ratio of 1:2 and surfactant at a surfactant-to-CNF weight ratio of 0.76:1. This mixture of CNFs, epoxy resin, surfactant, and acetone was then mixed in a planetary centrifuge mixer (Thinky-AR100) for three minutes. After this initial mixing, the mixture was sonicated for 4 hours in a bath sonicator. Sonication helps to break up CNF agglomerations. After the sonication process, a hotplate stirrer was used at 60 $^\circ$C for 24 hours in order to eliminate the acetone from the mixture via evaporation. Next, curing agent and air release
agent (BYK A-501) were added at weight ratios of 27:100 and 0.001:1, respectively. Lastly, the mixture was degassed for 30 minutes at room temperature before being poured in open rectangular molds and cured at 60 $^\circ$ for five hours. A total of five specimens were made at 0.25, 0.5, 1, 1.5, and 2.0 wt.\% CNFs.

\begin{figure*}[h]
\setlength{\fboxsep}{0pt}%
\setlength{\fboxrule}{0pt}%
\begin{center}
\includegraphics[width=\textwidth]{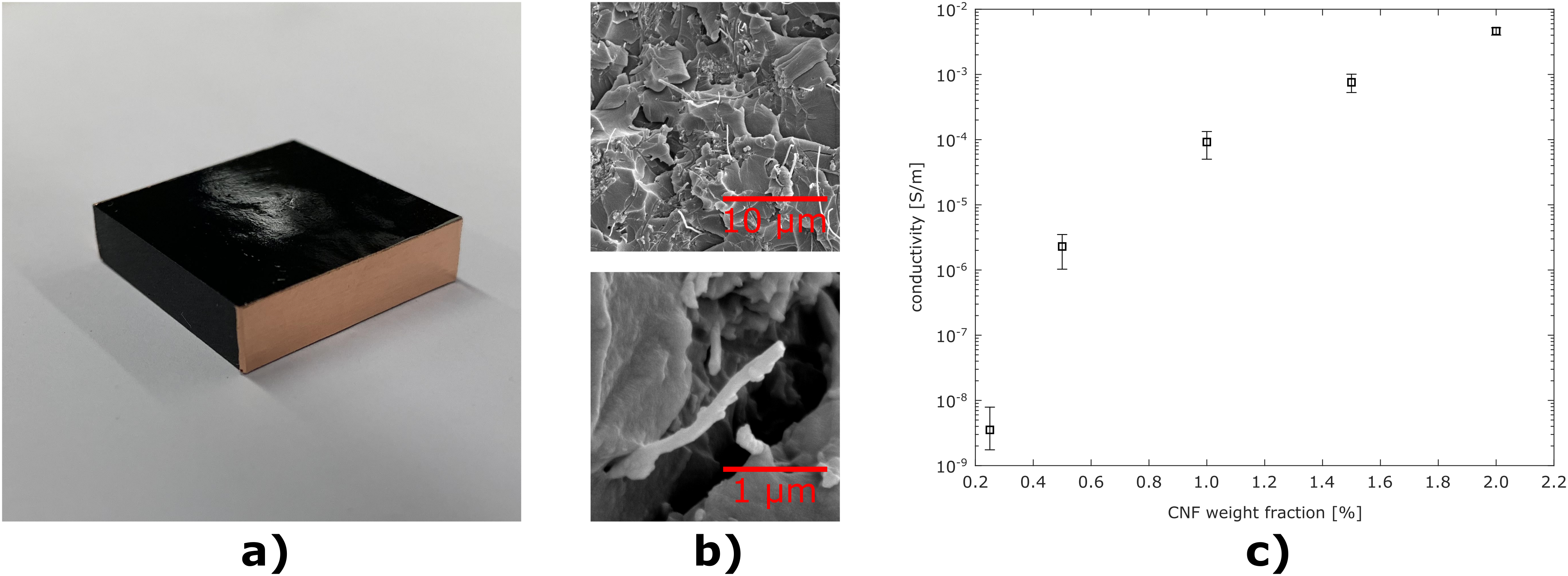}
\end{center}
\caption{Piezoresistive CNF/epoxy. a) Post-manufactured measurement specimen with electrodes applied. b) SEM images of CNF/epoxy fracture surface showing CNF dispersion above and close up of a single CNF protruding from the fracture surafce below. c) Plot of CNF/epoxy conductivity versus CNF weight fraction. Boxes indicate mean conductivity and error bars indicate standard deviation.} \label{Fig:cnf_epoxy_specimen_sem_plot}
\end{figure*}

Post-cured CNF/epoxy specimens were then cut into smaller pieces measuring 1'' $\times$ 1'' $\times$ 0.25'' via a water-cooled tile saw. These smaller pieces are referred to as measurement specimens. This resulted in a total of eight measurement specimens per CNF weight fraction. Electrodes were attached using a combination of conductive silver paste and copper tape such that electrical resistance could be measured. Post-manufactured and instrumented specimens and scanning electron microscope (SEM) images of the underlying CNF network are shown in Figure \ref{Fig:cnf_epoxy_specimen_sem_plot}. From the measured resistance and specimen dimensions, the electrical conductivity was calculated. Conductivity versus CNF weight fraction data is also shown in Figure \ref{Fig:cnf_epoxy_specimen_sem_plot}. This experimentally determined conductivity was used to calibrate the EIT model that is used to train the networks as described later.

\section{Electrical impedance tomography}

EIT is a process by which the internal conductivity of a domain is mapped. EIT has two important parts -- i) the forward problem by which the voltage-current relationship in the domain is modeled and ii) the inverse problem which seeks to spatially map the internal conductivity distribution from observations of the voltage-current relationship. These parts are both described in the fourthcoming sub-sections. 

\subsection{EIT forward problem}
The goal of the EIT forward problem is to solve for electrode voltages for a prescribed current injection magnitude and conductivity distribution. This process is governed by Laplace's equation in the absence of internal current sources as is shown in equation (\ref{laplace_eqn}) where $\sigma$ is the conductivity distribution and $\phi$ is the electric potential.


\begin{equation}\label{laplace_eqn}
    \boldsymbol{\nabla} \cdot \sigma \boldsymbol{\nabla} \phi=0
\end{equation}

In EIT, the complete electrode model (CEM) is used as boundary conditions on Laplace's equation. The CEM boundary conditions account for the finite size of practical electrodes by adding an additional degree of freedom for the electrode voltage. As shown in equation (\ref{CEM_cont_imp}), the CEM also accounts for a voltage drop between the domain and electrodes due to contact impedance. Charge conservation is enforced by equation (\ref{cons_charge}) by requiring that the total current entering the domain is equal to the total current leaving the domain. In these equations, $z_l$ represents the contact impedance between the $l$th electrode and the domain, $V_l$ is the voltage of the $l$th measurement electrode, the total number of electrodes is denoted by $L$, $E_l$ represents the size of the $l$th electrode, and $\boldsymbol{n}$ represents a normal vector pointing outward from the domain boundary.



\begin{equation}\label{CEM_cont_imp}
    z_{l} \sigma \nabla \phi \cdot \boldsymbol{n}=V_{l}-\phi
\end{equation}

\begin{equation}\label{cons_charge}
    \sum_{l=1}^{L} \int_{E_{l}} \sigma \boldsymbol{\nabla} \phi \cdot \boldsymbol{n} \; \mathrm{d} S_{l}=0
\end{equation}


Equations (\ref{laplace_eqn})-(\ref{cons_charge}) are most easily solved by discretization via the finite element method. This is shown in equation (\ref{FEM_CEM}). In these equations, $\boldsymbol{\Phi}$ is a vector representing electric potential in the domain, $V$ is a vector of electrode voltages, and the vector $I$ represents electrode currents. $\boldsymbol{A}_M$ is the stiffness matrix for steady-state diffusion and the remaining matrices are formed as shown as follows where $w_i$ is the $i$th finite element interpolation function. This work uses linear triangle elements.

\begin{equation}\label{FEM_CEM}
    \begin{bmatrix}
        \boldsymbol{A}_M+\boldsymbol{A}_Z & \boldsymbol{A}_W \\
        \boldsymbol{A}_W^T & \boldsymbol{A}_D
    \end{bmatrix}
    \begin{bmatrix}
        \boldsymbol{\Phi} \\
        \boldsymbol{V}
    \end{bmatrix}
    =
    \begin{bmatrix}
        0 \\
        \boldsymbol{I}
    \end{bmatrix}
\end{equation}

\begin{equation}\label{AZ_matrix}
    A_{Z\;ij}=\sum_{l=1}^{L} \int_{S_l} \frac{1}{z_l}w_i w_j\;\text{d}S_l
\end{equation}

\begin{equation}\label{AW_matrix}
    A_{W\;li}=-\int_{S_l} \frac{1}{z_l}w_i\;\text{d}S_l
\end{equation}

\begin{equation}\label{AD_matrix}
    A_D=\text{diag}\left(\frac{E_l}{z_l}\right)
\end{equation}


\subsection{EIT inverse problem}

The EIT inverse problem tries to spatially map the conductivity distribution of the domain for a given set of boundary voltage observations. This can be done by updating the conductivity distribution of the forward model until the model-predicted voltages match the experimentally collected voltages. However, recovering the conductivity distribution in this way is often impractical because the EIT inverse problem is ill-posed and therefore extremely susceptible to factors such as model-to-experiment electrode misplacement, geometry deviations, and noise. Therefore, difference imaging is most often used wherein the conductivity change between two states is sought (i.e. the change in conductivity before damage and after damage). Herein, we use a one-step linearization approach by linearizing the damage-induced conductivity change about an initial estimate of the background or unperturbed conductivity distribution. In order to do this, define the experimentally observed change in boundary voltage as shown next in equation (\ref{y_eqn}).

\begin{equation}\label{y_eqn}
    \delta\boldsymbol{V}=\boldsymbol{V}_m\left(t_2\right)-\boldsymbol{V}_m\left(t_1\right)
\end{equation}

Above, the $t_1$ and $t_2$ arguments indicate that boundary voltages were measured before and after the damage event, respectively. We can likewise define this difference via the forward problem as $\boldsymbol{W}=\boldsymbol{F}\left(\boldsymbol{\sigma}+\delta\boldsymbol{\sigma}\right)-\boldsymbol{F}\left(\boldsymbol{\sigma}\right)$ where $\boldsymbol{F}\left(\cdot\right)$ is the output of the forward problem evaluated at the conductivity in its argument and $\delta\boldsymbol{\sigma}$ is the damage-induced conductivity change which we seek. To find $\delta\boldsymbol{\sigma}$, we linearize $\boldsymbol{F}\left(\boldsymbol{\sigma}+\delta\boldsymbol{\sigma}\right)$ about  the unperturbed background conductivity, $\sigma$, via a truncated Taylor series as shown below in equation (\ref{linearization}).

\begin{equation}\label{linearization}
    \boldsymbol{F}\left(\boldsymbol{\sigma}+\delta\boldsymbol{\sigma}\right)\approx\boldsymbol{F}\left(\boldsymbol{\sigma}\right)+\frac{\partial\boldsymbol{F}\left(\boldsymbol{\sigma}\right)}{\partial\boldsymbol{\sigma}}\delta\boldsymbol{\sigma}
\end{equation}

To actually solve for the conductivity change, we form a minimization problem in which we seek to minimize the difference between $\delta\boldsymbol{V}$ and the linearized form of $\boldsymbol{W}$ as shown in equation (\ref{eit_min}).

\begin{equation}\label{eit_min}
    \delta\boldsymbol{\sigma}^*=\argmin_{\delta\boldsymbol{\sigma}\leq 0}\left(\left\|\boldsymbol{J}\delta\boldsymbol{\sigma-\delta\boldsymbol{V}}\right\|_m^m+\alpha\left\|\boldsymbol{R}\left(\delta\boldsymbol{\sigma}\right)\right\|_n^n\right)
\end{equation}

Above in equation (\ref{eit_min}), $m$ and $n$ represent the $m$ or $n$th vector norm to the $m$ or $n$th power. Herein, we use $m=n=2$, but interested readers are directed to work by Tallman and Hernandez \cite{tallman2017effect} for a discussion on other norms. Note also that a regularization term, $\boldsymbol{R}\left(\delta\boldsymbol{\sigma}\right)$, has been incorporated into the minimization. This is necessary because of the ill-posedness of the EIT inverse problem. A discretized approximation of the Laplace operator is used for regularizing the inverse problem in this work due to its ability to filter out highly oscillatory noise artifacts. The extent of regularization is controlled by the scalar hyper-paremeter, $\alpha$. Lastly, note that the minimization has been constrained such that $\delta\boldsymbol{\sigma }\leq 0$. This is based on the physical realization that damage will cause a complete cessation or loss of conductivity. For more detailed solution strategies of equation (\ref{eit_min}), interested readers are directed to the engaging work of Smyl et al. \cite{smyl2020overview}

\section{Problem setting}
In this study, the goal is to detect the damage status on the self-sensing material specimen by analyzing the EIT-generated conductivity change distribution data with machine learning methods. In this preliminary, proof-of-concept study, we consider only simple through-hole damages. This task is divided into three sub-tasks: predicting the number, the radius, and the center position of the holes.
For EIT images, we use convolutional neural networks (CNN) to predict the number of holes, the radius of holes, and use K-means to predict the center position of holes. For conductivity change vectors, we use fully connected neural networks (FCNN) to detect the damage status.

 For the problem of predicting the number of holes, the principal idea is to build a model that can be used to predict the number of holes given the EIT generated conductivity change data. Since in our dataset, the number of holes on a specimen is between 1 to 3, this problem can be seen as a multi-class classification problem, the class 1, class 2, and class 3 represent that this specimen has 1, 2, and 3 holes, respectively. 
With the EIT images as input, we can train a CNN to predict the number of holes. On the other hand, with the conductivity change vectors as input, the FCNN is trained as a classification model to predict the number of holes.
 
 In order to detect the degree of damage on the material specimens, the radius of holes is needed to be identified. A generic model is created to predict the radius of holes on material specimens given the EIT-generated conductivity change data. Since the distribution of radius is continuous, this task can be seen as a regression problem. With the EIT images as input, we can train a CNN to predict the radius of holes. On the other hand, with the conductivity change vectors as input, the FCNN is trained as a regression model to predict the radius of holes.

In order to localize the damage, we need to predict the center position of holes. From the EIT images, a hole can be seen as a cluster of pixels; therefore, the clustering algorithms can be applied to predict the center position of holes. Unlike the neural networks we introduced above, the clustering algorithms are unsupervised machine learning algorithms, so we do not need to provide an annotation of samples to the clustering algorithms. In our study, the K-means algorithm is used and the coordinates of the pixels that form the damaged domain on the image are fed to the K-means, K-means divides the coordinates into K clusters, K is the number of holes, and K-means can also output the centroids for the K clusters, we can approximately see the K centroids as the centers of holes. Using the conductivity change vectors as input, this problem can be treated as a regression problem, so we apply the FCNN as a regression model to predict the center position of holes with the vectors.
\section{Dataset}
\begin{figure}[h]
    \centering
    \includegraphics[width=8cm]{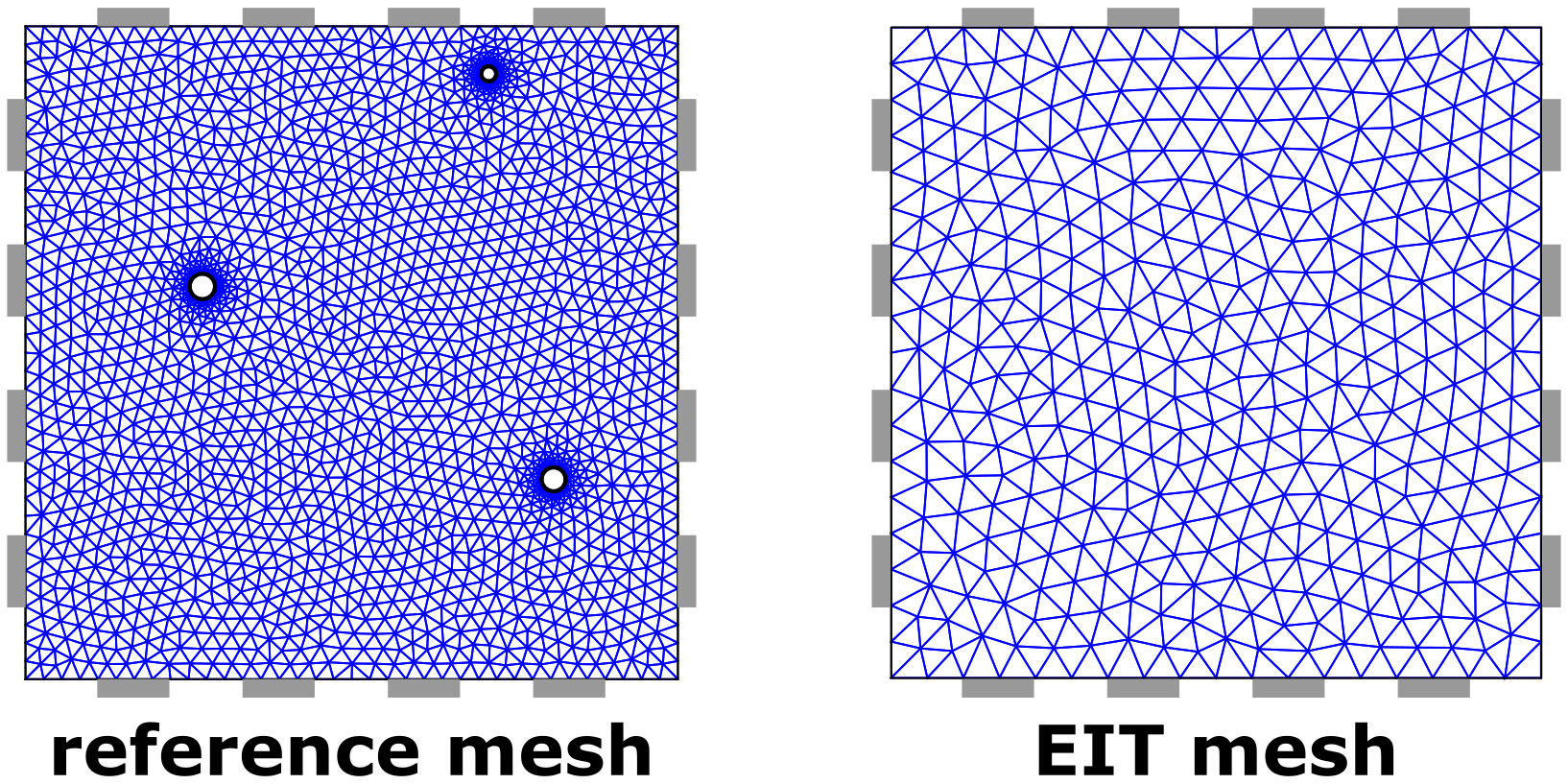}
    \caption{Representative damaged reference mesh (note that damage is randomly generated such that no two reference meshes are the same) and the EIT reconstruction mesh. Electrodes are shown along the periphery of the domains in gray.}
    \label{Fig:meshes}
\end{figure}

A large amount of data is required in our study, and all data used in our study is generated by an experimentally calibrated EIT simulation program. In these simulations, an adaptive mesh routine is used to simulate a random number of through holes within the specimen. Holes are simulated as voids in the mesh. The simulated specimen used in this work measured 0.9 m $\times$ 0.9 m and has a background conductivity equal to the experimentally measured mean conductivity of CNF/epoxy at 1.0 wt.\% CNFs, 9.16$\times$10\textsuperscript{-5} S/m. A total of 16 electrodes, each measuring $1/9$th of the domain width, were simulated using the CEM as described previously. These electrodes can be seen in Fig.\ref{Fig:meshes}. An adjacent injection and measurement scheme was used.

All damage in our study is circular and each specimen has at least 1 hole and at most 3 holes. For the balance of data, the numbers of generated samples with 1, 2, and 3 holes are the same. Each hole has a random radius between 0.03 m and 0.05 m. The EIT routine described previously is then used to reproduce the conductivity change distribution of the mesh with holes onto a different, hole-free mesh thereby avoiding a so-called `inverse crime.' Specifically, two references meshes are used for the difference imaging process -- a hole-free reference mesh and a reference mesh with 1-3 randomly generated holes such that a voltage change vector, $\delta V$, due to the holes can be formed. Further, the reference meshes are much more refined than the EIT reconstruction mesh. The reference meshes have on the order of 3,500 triangular elements (the exact number of elements depends on the number and size of the randomly generated through holes). The finite element mesh used for the EIT reconstruction, on the other hand, always has 772 triangular elements. Because finite element simulations are well known to converge with increasing number elements, utilizing a much coarser EIT mesh for the conductivity mapping ensures that intrinsic modeling errors exist between the reconstruction mesh and the simulated experimental data (i.e. the reference mesh) as would be expected for actual experimental data. A representative reference mesh with three holes and the EIT reconstruction mesh are shown in Fig.\ref{Fig:meshes}.To more faithfully reproduce experimental conditions, voltage measurements were also contaminated with noise at a signal-to-noise ratio of 85 dB. These simulation parameters (i.e. EIT mesh refinement, signal-to-noise ratio, choice of regularization for EIT, etc.) were selected based on our previous experience with using EIT for imaging of CNF-based polymeric nanocomposites.\cite{hassan2019failure,tallman2015tactile,tallman2014damage}


EIT generates two categories of data, conductivity change distribution vectors and images constructed from the vectors. For the vector, its length is $772$, which equals to the number of cells in the EIT finite element mesh. Each element in the vector represents the conductivity change value of the corresponding cell. The images are stored as 538-by-538-by-3 data matrix that records the red, green, and blue color components for each individual pixel. Both the width and height of the image is 538 pixels. 
Both vector-based and image-based data represent the conductivity change distribution through a specimen and are considered as the inputs of our machine learning models in this study. 



\section{Methodology}
\subsection{EIT image-based machine learning approach}

\subsubsection{Convolutional neural networks}

CNNs are a variant of neural networks that are frequently applied to image analysis. In this section, the most important theories and concepts of CNNs are described.

\textbf{Convolutional layer}

The convolutional layer is essential in CNN since it is used to detect features and has several kernels with learnable weights.\cite{cha2018autonomous} Kernels work as filters and a kernel is a matrix of integers. Each kernel provides a measure for how close a patch of input resembles a feature. A kernel slides over the complete image and dot product is taken between the kernel and a patch of the image. The greater the result, the closer the patch of the image resembles the feature. The computed dot product values corresponding to the channels of each kernel are summed up with a bias to produce the results of each kernel. These results form the spatial feature maps of the convolutional layer.


\textbf{Pooling layer}

The pooling layer is where the features extracted by the convolutional layer are selected for downsampling, which reduces the computational cost. The function of the pooling layer is straightforward; for example, the max pooling layer extracts the maximum in the feature maps; the average pooling layer extracts the averages in the feature maps.\cite{lee2018image} 

\textbf{Fully connected layer}

The fully connected layer is an essential component of CNNs because it connects to all nodes in the previous layer. It takes the outputs of the previous layer, performs dot products between the weights of each node and the inputs, and adds a bias in each node.\cite{schmidhuber2015deep}

\textbf{Softmax function}

The softmax function is an activation function usually used in the classification model to predict the class of the input. The softmax function takes the output from the previous layer, estimates the probabilities for each class, and predicts the class that has the greatest probability as the prediction of the classification model.\cite{kim2014convolutional}

\subsubsection{K-means}

The K-means algorithm is a clustering algorithm that is capable of partitioning a dataset into K distinct clusters\cite{krishna1999genetic} -- these clusters should not overlap each other such that each data point in the dataset belongs to only one cluster. K-means tries to make the data points inside a cluster as close as possible while keeping the data points in different clusters as far as possible. It randomly selects K data points as initial centroids then the sum of the squared distance between data points and all centroids are calculated. Each data point is assigned to the closest centroid, then centroids for the clusters are calculated by taking the average of the all data points that belong to each cluster. These steps are iterated until the centroids cannot be changed anymore.

\subsubsection{Data Preprocessing}
\begin{figure*}[h]
\setlength{\fboxsep}{0pt}%
\setlength{\fboxrule}{0pt}%
\begin{center}
\includegraphics[width=\textwidth]{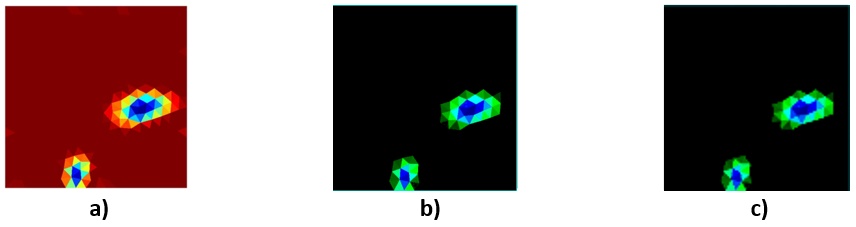}
\end{center}
\caption{Preprocessing of EIT images used for image-based training. a) An original EIT image. b) The image after being removed the color components. c) The image after being resized.} 
\label{Fig:img_preprocess}
\end{figure*}

Noise is a common feature of EIT images, which influences the performance of our model. In order to mitigate this, the red color components of pixels are removed, the green and blue components whose value are smaller than 50 are also removed. In order to reduce the dimension of input and accelerate the model training process, the input images are resized to 100 pixels $\times$ 100 pixels by applying interpolation. Pixel values of images are normalized by dividing all pixel values by 255 in order to improve the performance of the model and speed up the training process. The preprocessing steps are shown in Fig.\ref{Fig:img_preprocess}. 

\subsubsection{Prediction of the number of holes and the radius of holes}

\begin{figure*}[h]
\setlength{\fboxsep}{0pt}%
\setlength{\fboxrule}{0pt}%
\begin{center}
\includegraphics[width=\textwidth]{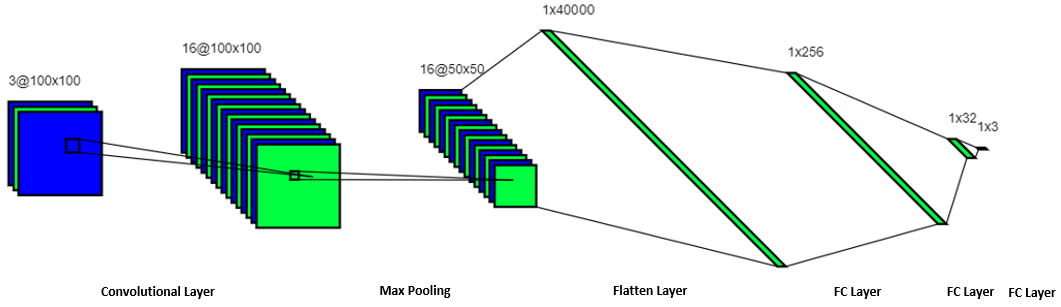}
\end{center}
\caption{The structure of the EIT image-based CNN for predicting the number of holes. 1 Convolutional layer, 2 Max pooling, 3 Flatten layer, 4 Fully connected layer, 5 Fully connected layer, 6 Fully connected layer + Softmax.} 
\label{Fig:CNN_num_structure_1}
\end{figure*}

Two CNNs are used to predict the number and radius of holes from EIT images. For the CNN used to predict the number of holes, the first layer is the convolution layer that has 16 kernels with a kernel size of 3-by-3. The second layer is a max pooling layer used to reduce the complexity of computation by reducing the dimensionality of the input. The third layer is a flatten layer that flattens the input to a 1-dimensional array. The fourth layer is a fully connected layer that has 256 nodes to capture all the information contained in the input of this layer. The fifth layer is a fully connected layer with 32 nodes. An activation function called ReLU is applied in nodes in the convolutional layer and the first and second fully connected layer. Lastly, the output layer is a fully connected layer that has 3 nodes with a softmax activation function to work as a classifier. Fig.\ref{Fig:CNN_num_structure_1} presents the details of the structure of the CNN clearly.






The structure of the CNN for predicting the radius of holes is similar to that for predicting the number of holes, the only difference is that the output layer has no activation function since the convolutional neural network is a regression model.

\subsubsection{Prediction of the center position of holes}

K-means alogirthm is employed to predict the center position of holes. The first step of using the K-means algorithm is specifying the number of clusters, $K$, which is the number of holes on the image in our study. A neural network model is employed first to predict the number of holes accurately as the value of K. After the EIT images are preprocessed, only the pixels that form the damage domain are reserved, and the coordinates of these pixels can be determined. The K-means algorithm with the coordinates is employed to predict the center position for the $K$ number of holes.




\subsection{Conductivity change vector-based machine learning approach}
\subsubsection{Data Preprocessing}

The conductivity change vector is normalized to a unit vector by dividing every element by the norm of the vector. This normalization procedure can speed up the gradient descent process and improve the performance and stability of our model.

\subsubsection{Prediction of the number, radius and center position of holes}

\begin{figure*}[h]
\setlength{\fboxsep}{0pt}%
\setlength{\fboxrule}{0pt}%
\begin{center}
\includegraphics[width=\textwidth]{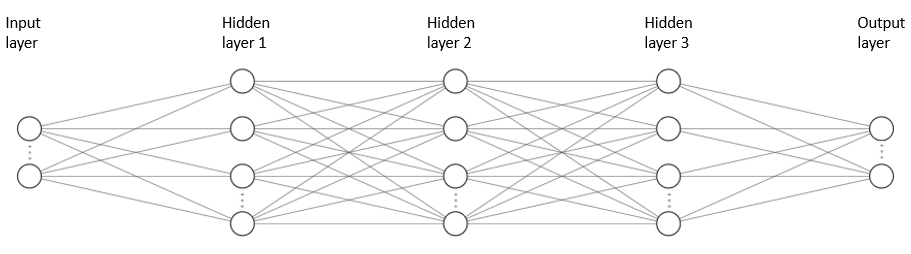}
\end{center}
\caption{The structure of the conductivity change vector-based FCNN for predicting the number of holes.} 
\label{Fig:FCNN_num_structure}
\end{figure*}

Three FCNNs are employed to predict the number, the radius, and the center position of holes separately. First, a FCNN is built as a multi-class classification model to predict the number of holes in three classes (i.e.e classes with 1, 2, and 3 holes) from the conductivity change vectors. The proposed network is composed of 1 input layer, 3 hidden layers, and 1 output layer. The input layer contains 772 nodes. Each of them represents an element in the conductivity change vector (recall that the EIT reconstruction mesh has 772 elements). The number of nodes in the first hidden layer is 256, which is enough to understand all the information contained in the input layer. There are 64 and 16 nodes in the next two hidden layers. The ReLU activation function is applied in nodes in hidden layers. In the output layer, there are three nodes and the activation function is again a softmax function since our model is a three-classes classifier. Fig.\ref{Fig:FCNN_num_structure} presents the structure of the FCNN.

Second, a FCNN was built as a regression model to predict the radius of holes with the conductivity change vectors. The structure of the FCNN is almost the same as that for predicting the number of holes except that the third hidden layer has 32 neurons and no activation function is applied on the output layer in this network since the model is a regression model.






Third, a FCNN is constructed as a regression model to predict the center position of holes from the conductivity change vectors. This FCNN is similar to the one for predicting the number of holes except that there are 128 and 32 nodes in the second and third hidden layers and there are six neurons in the output layer with no activation function. 





\section{Results using the machine learning approaches with experimentally calibrated simulation data}

\subsection{Results obtained from the EIT image-based machine learning approach}

\begin{figure}[htb]
    \centering
    \includegraphics[width=8cm]{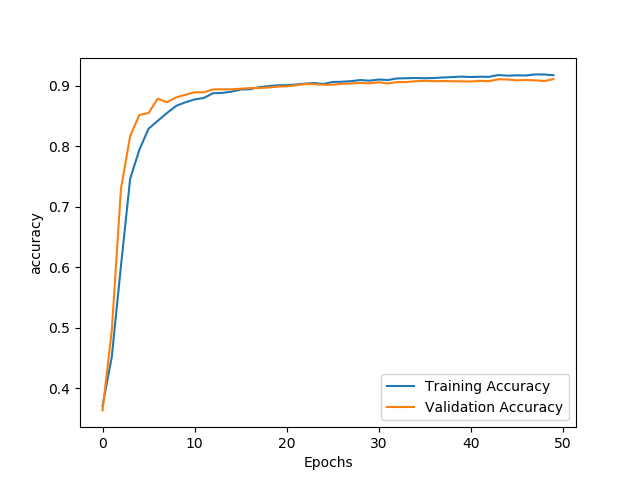}
    \caption{Training process of EIT image-based CNN for predicting the number of holes.}
    \label{Fig:CNN_num_acc}
\end{figure}

The first experiment is conducted using the previously proposed CNN for predicting the number of holes. The network is trained and tested with EIT images. $22,950$ EIT images are used to train and validate the CNN. In the training process, 50 epochs and batch size 128 have been used when feeding in the training samples. After training, the model's performance on $4,050$ testing images gives the accuracy of 0.911. Fig.\ref{Fig:CNN_num_acc} shows the training accuracy and the validation accuracy plots against epochs. From Fig.\ref{Fig:CNN_num_acc}, we can see the training error and validation error converge after 50 epochs. 



\begin{figure}[htb]
    \centering
    \includegraphics[width=8cm]{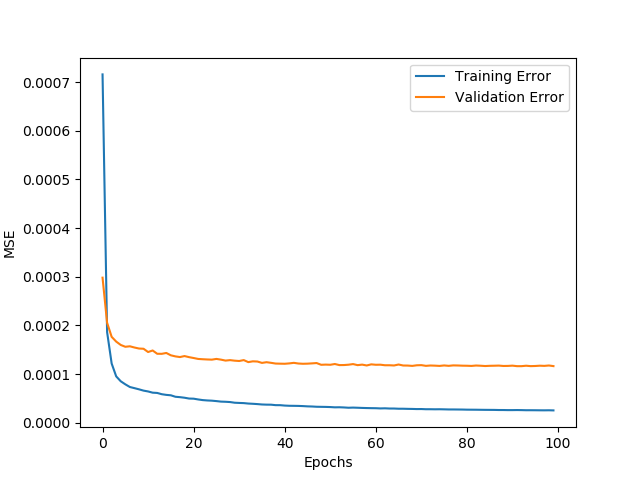}
    \caption{Training process of EIT image-based CNN for predicting the radius of holes.}
    \label{Fig:CNN_rad_loss}
\end{figure}

The second experiment is conducted using the previously proposed CNN for predicting the radius of holes. $22,950$ EIT images are used to train the neural network. After the training, the model is tested on $4,050$ images. The performance using mean squared error (MSE) gives a testing error of $1.16 \times 10^{-4}$. By comparing the predicted output and the annotation, the average difference between the predicted and true radius is $0.00511$ m.  Fig.\ref{Fig:CNN_rad_loss} shows the training process. From Fig.\ref{Fig:CNN_rad_loss}, we can see the training error and validation error converge -- the validation error is relatively large compared to the training error.      

 The third experiment is conducted using the K-means model with $27,000$ images to predict the center position of holes. By comparing the predicted output and annotations, the average distance between predicted and true center position is $0.0172$ m. 

\subsection{Results obtained from the conductivity change vector-based machine learning approach}

\begin{figure}[h]
    \centering
    \includegraphics[width=8cm]{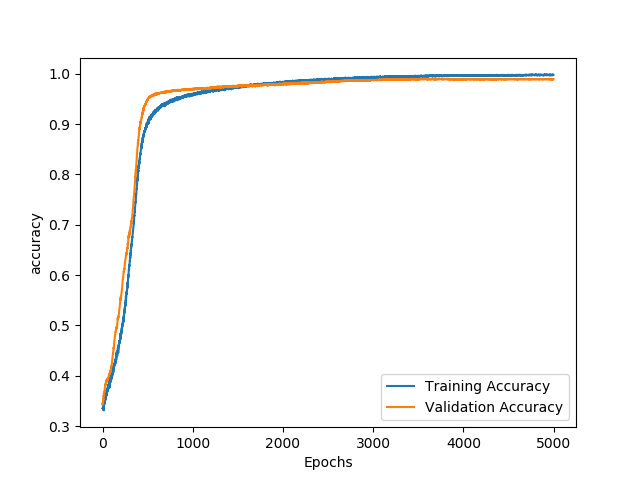}
    \caption{Training process of the conductivity change vector-based FCNN for predicting the number of holes}
    \label{Fig:EXP_FCNN_num_acc}
\end{figure}

The first experiment is conducted using the previously proposed FCNN for predicting the number of holes. $21,600$ conductivity change vectors are used to train and validate the fully connected neural network. $5,000$ epochs and batch size 128 have been applied in the training process. After the training, the model’s performance on $5,400$ testing  samples is evaluated with  an  accuracy  of  0.992. Fig.\ref{Fig:EXP_FCNN_num_acc} shows the training accuracy and the validation accuracy as a function of epochs. From this result, we can see a clear improvement in accuracy compared to the result of using EIT images as the inputs. 

\begin{figure}[h]
    \centering
    \includegraphics[width=8cm]{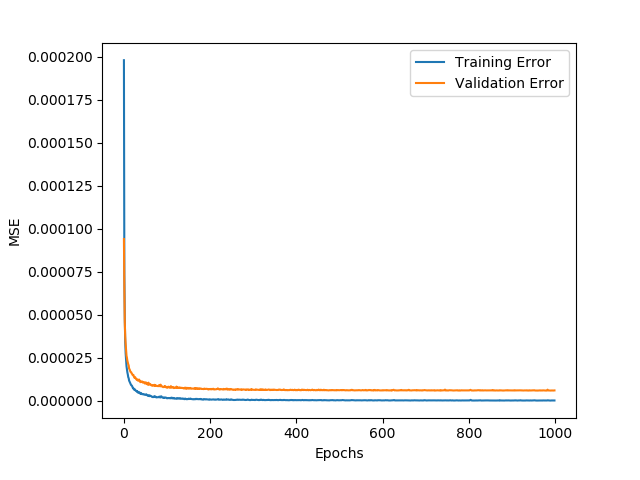}
    \caption{Training process of the conductivity change vector-based FCNN for predicting the radius of holes.}
    \label{Fig:EXP_FCNN_rad_loss_1}
\end{figure}

The second experiment is conducted using the previously proposed FCNN for predicting the radius of holes. 
$21,600$ conductivity change vectors are used to train and validate the FCNN. $1,000$ epochs and batch size $256$ have been applied in the training process. After the training, the model’s performance on $5,400$ testing  samples using MSE is evaluated with the testing error of $6.11 \times 10^{-6}$. By comparing the predicted output and the annotation, the average difference between the predicted and true radius is $0.000987$ m. Fig.\ref{Fig:EXP_FCNN_rad_loss_1} shows both the training error and validation error decrease as increasing the epochs.

\begin{figure}[h]
    \centering
    \includegraphics[width=8cm]{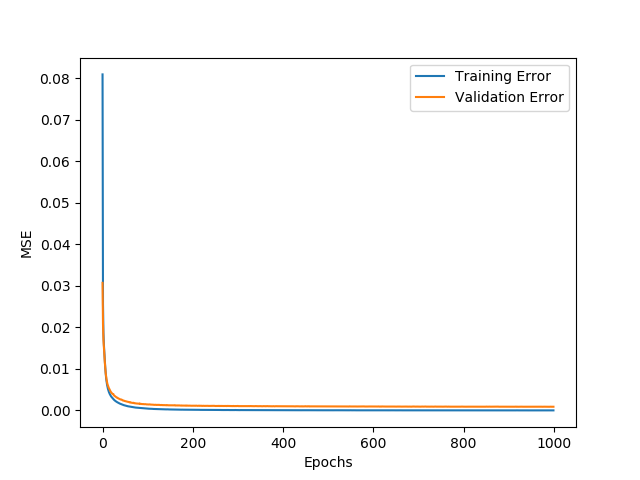}
    \caption{Training process of the conductivity change vector-based FCNN for predicting the center position of holes.}
    \label{Fig:EXP_FCNN_cnt_loss_1}
\end{figure}

The third experiment is to use the proposed FCNN for predicting the center position of holes.
$21,600$ conductivity change vectors are used to train and validate the FCNN, $1,000$ epochs and batch size 64 have been applied in the training process. After the training, the model’s performance on $5,400$ testing  samples is evaluated using MSE with the testing error is $3.21 \times 10^{-4}$. By comparing the predicted output and the annotation, the average difference between the predicted and true center position is 0.00808m. Fig.\ref{Fig:EXP_FCNN_cnt_loss_1} shows the training error and validation error decrease as increasing epochs during the training process of the neural network.

\subsection{Comparison of results with EIT images and conductivity change vectors as input}
\begin{figure*}[htb]
    \setlength{\fboxsep}{0pt}%
    \setlength{\fboxrule}{0pt}%
    \begin{center}
    \includegraphics[width=\textwidth]{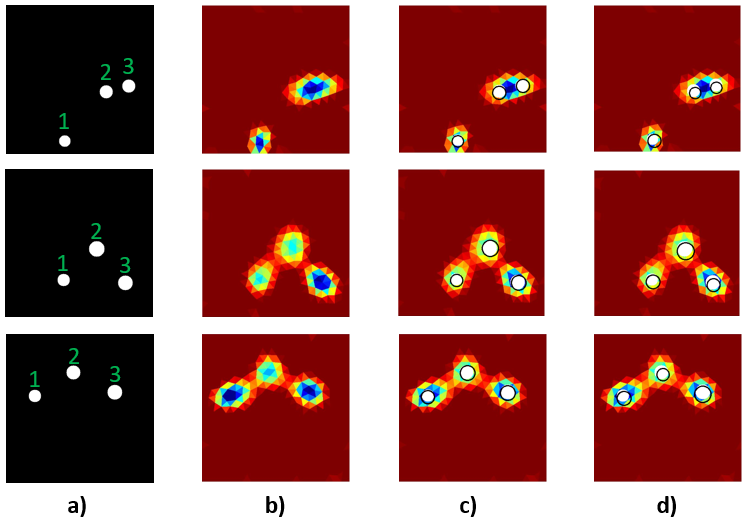}
    \end{center}
    \caption{Three representative examples of the performance of damage detection using machine learning approaches. a) Real damage status. b) Image generated by EIT. c) Damage predicted by the conductivity change vector-based machine learning approaches. d) Damage predicted by the EIT image-based machine learning approaches. In c) and d), the white filled circles are the real damages and the black circles are the predicted damages. Note again the top row where the proximity of the holes causes EIT to produce only a single region of conductivity change encompassing both holes. The method herein proposed, however, can adeptly recognize this as due to two distinct holes.}
    \label{Fig:compare_three_img}
\end{figure*}

According to our experiments, machine learning methods, especially neural networks, can be built for the damage prediction and quantification from EIT-generated conductivity change data. In our study, we use EIT images and conductivity change vectors as input separately to predict the number, the radius and the center position of holes. The comparison between the performance of models with different inputs is shown in the Table \ref{Table:The detailed specifications of NN1} and a few representative images are shown in Fig.\ref{Fig:compare_three_img}. From these results, we can clearly see that the conductivity change vector-based machine learning approach is a better option. The main reason may be that the process of constructing images from vectors causes the loss of information.

\section{Conclusions}
This study is one of the first attempts at combining neural networks with the EIT method for damage detection in self-sensing materials. And from the results presented herein, the proposed approach seems to have good performance. Compared to the traditional ways of using EIT for damage detection, our method not only detects the existence of damage but also gives a much more quantitative description of the damage status, including the number, radius, and center position of damage. Our results show that the machine learning approach can indeed predict the number of damages with up to 99.2\% accuracy, quantify damage size with respect to the averaged radius at an average of 2.46\% error, and quantify damage position with respect to the domain length at an average of 0.89\% error. And because all of this can be done in virtually real-time, this method seemingly has great potential for application to the field of SHM. However, utilizing the conductivity change vector directly considerably outperformed image recognition-based methods. This is likely a consequence of reducing the image quality prior to employing image recognition.

Encouraged by this success, several directions of future work are suggested. First, the size of damage studied in our research is not very small compared to the size of the specimen, particularly for damage detection thresholds of interest to practical SHM. Future work should therefore look to improve the sensitivity of our method for detecting smaller damage. Second, the number of holes in our study is limited between 1 to 3. Moving forward, we need to increase the number of holes that our method can handle. Third, in this proof-of-concept study, damage was limited to simple circular through-holes. Real structural damage is obviously much more complex. Consequently, future work should seek to generalize the shaping capabilities of the basic approach herein presented. Fourth, as described in section 5 Dataset, care was taken to ensure the simulation parameters are representative of real experimental data (i.e. by using a greatly refined reference mesh to ensure system modeling errors between the simulated data and the EIT mesh, including noise, etc.). However, the potential of the preliminary proof-of-concept results herein presented will need to ultimately be experimentally validated using real data. This will be especially important for understanding how modeling errors corrupt or bias the networks ability to predict actual damage. And fifth, the performance of neural network-aided damage characterization via EIT should be benchmarked against manual interpretation (i.e. EIT images interpreted by a person) in order to conclusively prove the advantage of machine learning over human inspection in this application.

\begin{table}[H]
  \caption{Comparison between the performance of models using EIT images and conductivity change vectors as input}
  \begin{center}
    \begin{tabular}{|>{\centering\arraybackslash}p{40pt}|>{\centering\arraybackslash}p{65pt}|>{\centering\arraybackslash}p{50pt}|>{\centering\arraybackslash}p{50pt}|}
      \hline
      \bf Prediction Goal & \bf Metrics& \bf EIT Images& \bf Conductivity Change Vectors\\
      \hline
      Number of holes& Testing accuracy& 0.911& 0.992\\
      \hline
      Hole radius& Average difference between predicted and true radius (m)& 0.00511& 0.000987\\
      \hline
      Center position& Average distance between predicted and true center positions (m)& 0.0172& 0.00808\\
      \hline
    \end{tabular}
  \end{center}
  \label{Table:The detailed specifications of NN1}
\end{table}

\section*{Conflicts of interest}
``There are no conflicts to declare.''

\section*{Acknowledgements}
We gratefully acknowledge the support from the National Science Foundation (DMS-1555072, DMS-1736364, CMMI-1634832, and CMMI-1560834), Brookhaven National Laboratory Subcontract 382247, ARO/MURI grant W911NF-15-1-0562 and Department of Energy DE-SC0021142.



\balance


\bibliography{rsc} 
\bibliographystyle{rsc} 

\end{document}